\documentclass[sigconf]{acmart} 
\usepackage{colortbl}
\usepackage{multirow}

\AtBeginDocument{%
  \providecommand\BibTeX{{%
    \normalfont B\kern-0.5em{\scshape i\kern-0.25em b}\kern-0.8em\TeX}}}

\setcopyright{acmlicensed}
\copyrightyear{2018}
\acmYear{2018}
\acmDOI{XXXXXXX.XXXXXXX}

\acmConference[Conference acronym 'XX]{Make sure to enter the correct
  conference title from your rights confirmation emai}{June 03--05,
  2018}{Woodstock, NY}
\acmBooktitle{Woodstock '18: ACM Symposium on Neural Gaze Detection,
 June 03--05, 2018, Woodstock, NY} 
\acmISBN{978-1-4503-XXXX-X/18/06}

\begin{document}

\title{MulliVC: Multi-lingual Voice Conversion With Cycle Consistency}


\author{Jiawei Huang}
\authornote{Both authors contributed equally to this research.}
\authornote{Interns at ByteDance.}
\email{huangjw@zju.edu.cn}
\affiliation{%
\country{China}
}

\author{Chen Zhang}
\authornotemark[1]
\email{zhangchen.0620@bytedance.com}
\affiliation{%
\country{China}
}

\author{Yi Ren}
\email{ren.yi@bytedance.com}
\affiliation{%
\country{China}
}

\author{Ziyue Jiang}
\email{ziyuejiang@zju.edu.cn}
\authornotemark[2]
\affiliation{%
\country{China}
}

\author{Zhenhui Ye}
\email{ziyuejiang@zju.edu.cn}
\authornotemark[2]
\affiliation{%
\country{China}
}

\author{Jinglin Liu}
\email{liu.jinglin@bytedance.com}
\affiliation{%
\country{China}
}
\author{Jinzheng He}
\authornotemark[2]
\email{jinzhenghe@zju.edu.cn}
\affiliation{%
\country{China}
}

\author{Xiang Yin}
\email{yinxiang.stephen@bytedance.com}
\affiliation{%
\country{China}
}

\author{Zhou Zhao}
\email{zhaozhou@zju.edu.cn}
\authornote{Corresponding author.}
\affiliation{%
\country{China}
}


\begin{abstract}
   Voice conversion aims to modify the source speaker's voice to resemble the target speaker while preserving the original speech content. Despite notable advancements in voice conversion these days, multi-lingual voice conversion (including both monolingual and cross-lingual scenarios) has yet to be extensively studied. It faces two main challenges: 1) the considerable variability in prosody and articulation habits across languages; and 2) the rarity of paired multi-lingual datasets from the same speaker. In this paper, we propose MulliVC, a novel voice conversion system that only converts timbre and keeps original content and source language prosody without multi-lingual paired data. Specifically, each training step of MulliVC contains three substeps: In step one the model is trained with monolingual speech data; then, steps two and three take inspiration from back translation, construct a cyclical process to disentangle the timbre and other information (content, prosody, and other language-related information) in the absence of multi-lingual data from the same speaker. Both objective and subjective results indicate that MulliVC significantly surpasses other methods in both monolingual and cross-lingual contexts, demonstrating the system's efficacy and the viability of the three-step approach with cycle consistency. Audio samples can be found on our demo page (\href{mullivc.github.io}{mullivc.github.io}).
\end{abstract}

\ccsdesc[500]{Computing methodologies~Natural language processing}
\ccsdesc[300]{Computing methodologies~Neural networks}
\ccsdesc[500]{Applied computing~Sound and music computing}

\keywords{voice conversion, multi-lingual voice conversion, speech disentanglement, cycle consistency}

\received{20 February 2007}
\received[revised]{12 March 2009}
\received[accepted]{5 June 2009}

\maketitle

\section{introduction}
These days, voice conversion (VC) has seen significant advancements owing to the emergence of various pretrained speech representations and major progress in speech synthesis models.
VC can be broadly classified into parallel and non-parallel systems~\cite{godoy2011voice} according to the type of training data. Considering the difficulty of getting parallel speech data~\cite{XVC-first} (source speaker and target speaker record the same speech content), non-parallel techniques are mainstream for voice conversion. Non-parallel techniques mainly focus on disentangling the content and speaker information from the source speech data and then reconstructing the speech~\cite{diffhiervc,ConsistencyVC,softvc,freevc,CyclePPG-XVC} using target speaker information. 
Some earlier work~\cite{CyclePPG-XVC,PPG-m2o,PPG-Xm2m} typically used pre-trained automatic speech recognition (ASR) models to extract Phoneme posterior-gram (PPG) as content features. Due to the emergence of large-scale pre-trained self-supervised learning (SSL) models such as Hubert~\cite{hubert}, WavLM~\cite{wavlm} and Wav2Vec~\cite{wav2vec}, recent voice conversion models~\cite{softvc,freevc,diffhiervc} tend to use SSL to extract content features and use speaker ID or speaker encoder to encode speaker information. However, the content features extracted by SSL models still contain the prosody and timbre information, leading to inadequate voice conversion\cite{freevc,nansy,contentvec}.

To bridge the gap between different languages, multi/cross-lingual voice conversion~\cite{XVC-ivec,PPG-Xm2m} has been developed, which is a special case in nonparallel systems and is more challenging. Given the high costs associated with gathering bilingual speaker datasets, current methods collect monolingual speaker datasets of different languages to train the models. The content and speaker information are sourced from the same language throughout the training phase; however, during inference, the disparity emerges as content and speaker information come from different languages. The inherent prosodic and pronunciation differences among various languages place traditional multi/cross-lingual VC methods into an out-of-domain context, resulting in generated speech with compromised intelligibility and speaker similarity.

In order to bolster the lingual generalization and the disentangle performance of the VC models, we propose MulliVC, a novel multi-lingual VC system that leverages cycle training strategy. Specifically, we divide each training iteration into three substeps. step 1 is the same as a traditional VC training step, we synthesize the speech  by using the content and timbre information from the same speaker. In step 2, we use the speech of two speakers who speak different languages as content and timbre inputs, constraining the outputs using timbre loss and asr loss. During step 3, we reconstruct the speech, preserving the timbre identified in step 2, by combining it with content from a different sentence by the same speaker, thus constituting a cross-lingual voice conversion cycle. 
Though we have no multi-lingual paired data, by strategically designing the information flow within the cycle comprising step 2 and step 3, we compel the model to exclusively learn timbre information from the timbre input while disregarding any extraneous information present in that input, which encourages the disentanglement of timbre and other information.
Furthermore, to improve the model's effectiveness in extracting timbre information, we introduce the fine-grained timbre conformer, designed to aid the model in capturing subtle aspects of timbre.
The experimental results denote that MulliVC outperforms baseline models in terms of both objective and subjective metrics and achieves substantial gains across monolingual and cross-lingual voice conversion scenarios.
\section{Related Works}
\subsection{Cross-lingual Voice Conversion}
Cross-lingual voice conversion aims to modify a source monolingual speaker's identity towards a target speaker who speaks another language while preserving the source linguistic content. It is more challenging than conventional monolingual voice conversion. ~\cite{XVC-ivec} proposes to use a bilingual Phonetic PosteriorGram for the content representation of speech, together with an averaging model~\cite{AMA} designed to synthesize the average speech that embodies the speakers present in the dataset. The averaging model serves as a generative model which is paired with an i-vector~\cite{AMA} seating adaptation step computed using a speaker verification formula~\cite{SV-formula}, to synthesize the target speaker's speech to achieve Cross-Language Voice Conversion (XVC). ~\cite{PPG-Xm2m} proposes to use a jointly trained speaker encoder instead of i-vector for better XVC results. 
FastSpeech-VC~\cite{FastSpeech-VC} pointed out that there are significant mismatches between phonetic sets and speech prosody of different languages, and PPG alone does not preserve rhythms well, for which they introduced normalized logarithm-scale fundamental frequency (Log-F0) to compensate for the prosodic mismatches. CyclePPG-XVC~\cite{CyclePPG-XVC} points out that the loss of spectral reconstruction optimized to match the identity of the target speaker causes the transformation model to capture the articulation of the target speech from a different language and the native pronunciation or articulation of the source speech cannot be preserved, making the intelligibility of the converted speech worse. For this reason, they introduced a cyclic loss on the PPG features to force the converted speech to carry the same linguistic content as the natural input speech. ConsistencyVC~\cite{ConsistencyVC} argues that some previous VC work used pre-trained speaker encoders in speaker classification tasks to obtain speaker embeddings, which are then used to guide speech synthesis. The main goal of the pre-trained speaker encoders is not speech synthesis, but speaker recognition. Therefore, this approach may miss emotional information in the reference speech. They use a jointly trained speaker encoder and after certain steps, this jointly trained speaker encoder is used to compute speaker consistency loss for improving speaker similarity and emotion similarity.

\subsection{Multi-lingual TTS}
Multi-lingual Text-To-Speech(TTS) is to synthesize speech in multiple languages, where the speaker's language can be the same or different from the target language. ~\cite{MTTS-liu} uses a Tacotron~\cite{tacotron} synthesizer with shared phonemes for inputs and a speaker encoder module to achieve multilingual TTS. They introduce tone/stress embeddings to represent tone and stress information for speech generation with native accents. To improve the speaker similarity between the synthesized speech and the recordings of the native speaker, ~\cite{MTTS-He} introduces multi-task learning and
speaker classifier joint training, they additionally add the speaker classification Cross Entropy loss and cross-lingual loss to the original loss. ~\cite{MTTS-lee} argues that the L2 (second-language) accents problem often occurs in cross-linguistic TTS and uses vowel space analysis, to study the L2 accents problem. They point out that the L2 accents of the parallel architecture (Glow-TTS)~\cite{glowtts} are less than the L2 accents of the autoregressive architecture (Tacotron). ~\cite{XTTS-cai} explores cross-lingual TTS in data-sufficient and low-resource scenarios. They propose that models that work well in data-sufficient scenarios do not perform well in low-resource scenarios for cross-language TTS. For this reason, they synthesize a pipeline that consists of a bilingual TTS system, a bottleneck feature extractor, a speaker embedding extractor, a multi-speaker voice conversion system, and a vocoder to achieve cross-language TTS. VALL-E X~\cite{vallex} uses a rule-based converter to convert the transcriptions to phoneme sequences and uses a neural codec encoder~\cite{encodec} to convert the speech into acoustic tokens. Then they concatenate the paired phoneme and acoustic token sequences of each language and train a multi-lingual conditional language model with a language ID module to alleviate accent problems. The generated acoustic tokens will be sent to the codec decoder to generate speech.

\subsection{Cycle/Back-Translation}
Back-translation~\cite{dual-backtrans,mono-backtrans} technique was first introduced in machine translation. It brings about the bridge between source and target languages by using a backward model that translates data from target to source. The (source and target) monolingual data is translated back and forth iteratively to progress the machine translation model in both directions. It is particularly effective in the case of missing data for parallel bilingual data. After that, some researchers introduced back-translation into the field of speech. In a low-resource scenario lacking text-to-speech alignment data, they use ASR models to generate pseudo-labels for speech. Then they use TTS models to regenerate speech with the transformed pseudo-labels, and the two models were jointly trained to achieve the training of an unsupervised TTS model~\cite{liu2022simple,unsupervisedtts,bagoftrick}. In the field of voice conversion, there is also some research with ideas similar to back-translation. CycleGAN-VC~\cite{cyclegan-vc} achieves one-to-one voice conversion without parallel data by jointly training two GAN models, one responsible for converting the speech of A to the speech of B timbre and the other vice versa. However, our approach is different to CycleGAN-VC, we use one single model to perform voice conversion between the two languages. In addition, our "back-translation" process maintains the timbre input as the same speaker, rather than keeping the speech content unchanged. We will discuss the detials in section~\ref{sec:train_strategy}.

\section{methods}

In this section, we will first describe the training pipeline overview of MulliVC. Next, we provide the cycle training strategy of MulliVC, which aims to improve the cross-lingual performance and timbre disentanglement of the model. Finally, we present the model architecture of MulliVC.

\subsection{Pipeline Overview}\label{sec:train_strategy}
\begin{figure*}[!t]
    \centering
    \includegraphics[width=\textwidth]{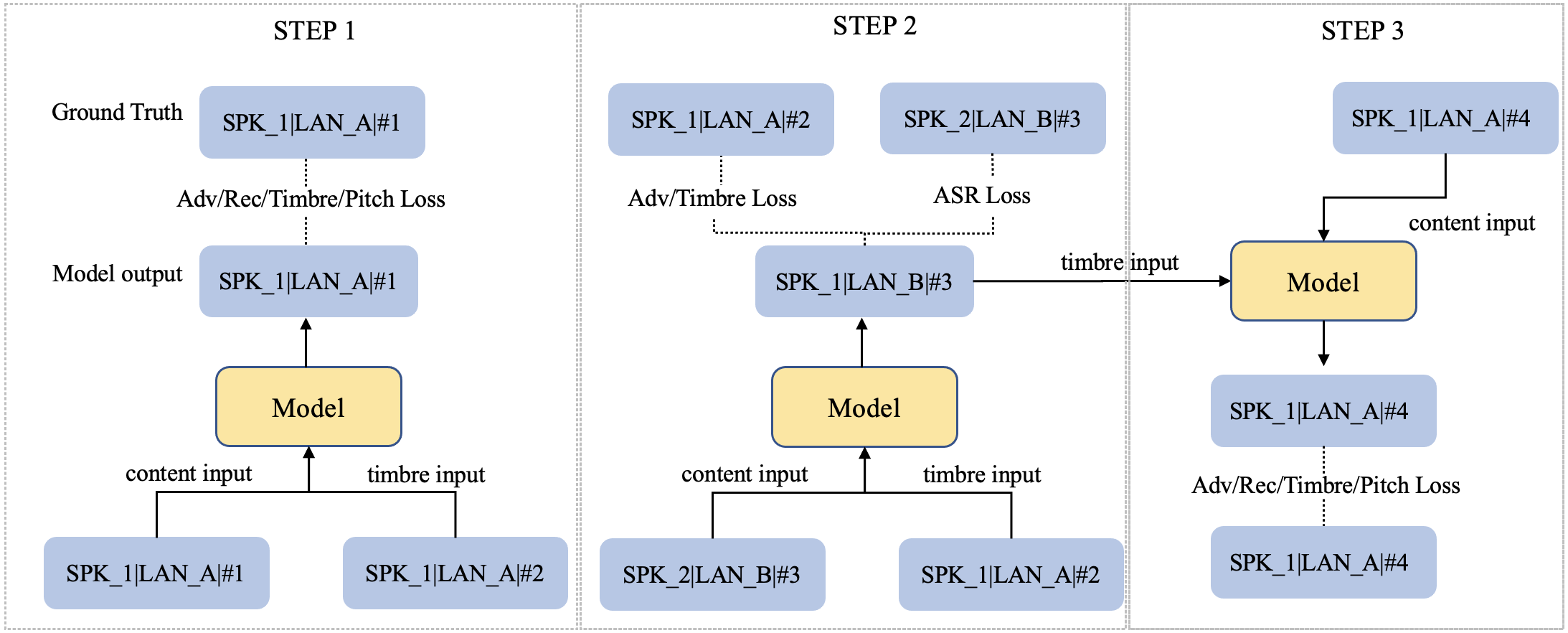}
    \caption{Training of MilliVC. SPK\_1|LAN\_A|\#2 denotes speech\#2 said by speaker 1 who speaks language A.} 
    \label{fig:train}
\end{figure*}

Obtaining data for the same speaker speaking multiple languages is a expensive and difficult task. Consequently, existing XVC methods often rely on combining monolingual speaker data to create a multilingual dataset. These methods aim to disentangle content and timbre information from speech and reconstruct speech using these two components. However, since the timbre and content information used during training belong to the same speaker's speech, existing models struggle to generate speech where the content information is from language A and the timbre information is from language B. As a result, suboptimal results are obtained. To address this limitation and fully leverage the potential of multilingual data, we propose a cycle training strategy.

Suppose we possess a large multilingual corpus consisting of two languages, namely language A and language B, with numerous speakers fluent in both languages. We randomly select two speakers for this illustration: Speaker 1, who speaks language A, and Speaker 2, who speaks language B. We represent the speech spoken by Speaker 1 in language A as SPK\_1|LAN\_A|\#1, where speech\#1 and speech\#2 refer to two distinct utterances.

As depicted in Figure \ref{fig:train}, each training step of our proposed MulliVC model consists of three substeps, and the losses from all three steps are summed up to perform a single model update. The training process of step 1 is similar to the previous works~\cite{diffhiervc,ConsistencyVC,freevc} in voice conversion, where the model takes the speech of the same person (take speaker 1 as an example in the figure) as both content input and timbre input to reconstruct the voice used as the content input, preserving the ability to transfer the timbre information when both inputs are in the same language. The loss of step 1 can be expressed as:
\begin{equation}
    \mathcal{L}=\lambda_1 \mathcal{L}_{Adv} + \lambda_2 \mathcal{L}_{Rec} + \lambda_3 \mathcal{L}_{Timbre}+ \lambda_4 \mathcal{L}_{Picth}
\end{equation}
\begin{equation}
    \mathcal{L}_{Timbre} = 1 - \frac{f_T(m_t)\cdot f_T(\hat{m_t})}{\lVert f_T(m_t)\rVert \cdot \lVert f_T(\hat{m_t})\rVert}
\end{equation}
Where $\lambda_{...}$ are weighting parameters, we set $\lambda_1=0.05,\lambda_2=1,\lambda_3=0.1,\lambda_4=1$ in our experiment. $\mathcal{L}_{Rec}=\lVert m_t - \hat{m_t} \rVert_2$ is the reconstruction loss, $\lVert \cdot \rVert_2$ is the L2 norm distance. $\mathcal{L}_{Adv}$ is the LSGAN-styled adversarial loss~\cite{lsgan} whose objective is to minimize the distribution distance between the generated mel-spectrograms $\hat{m_t}$ and the ground truth mel-spectrograms $m_t$ to avoid excessive smoothing problem. We adopt patch-based discriminator\cite{patch-gan} as our discriminator. $\mathcal{L}_{Picth} = \lVert f_{P1}(m_t) - f_{P1}(\hat{m_t}) \rVert_2$ is the pitch perceptual loss, $f_P$ is a pre-trained pitch predictor, we use the first layer embedding to calculate the perceptual loss. $\mathcal{L}_{Timbre}$ is the timbre loss, where $f_T$ is a pre-trained speaker verification(SV) model.

The content input and timbre input of step 2 and step 3 come from different languages, simulating a cross-language voice conversion scenario. The output of step 2 is used as the timbre input of step 3, and the two together form a cycle consistency loop, which will be detailed in the next section.

\begin{figure*}[!t]
    \centering
    \includegraphics[width=\textwidth]{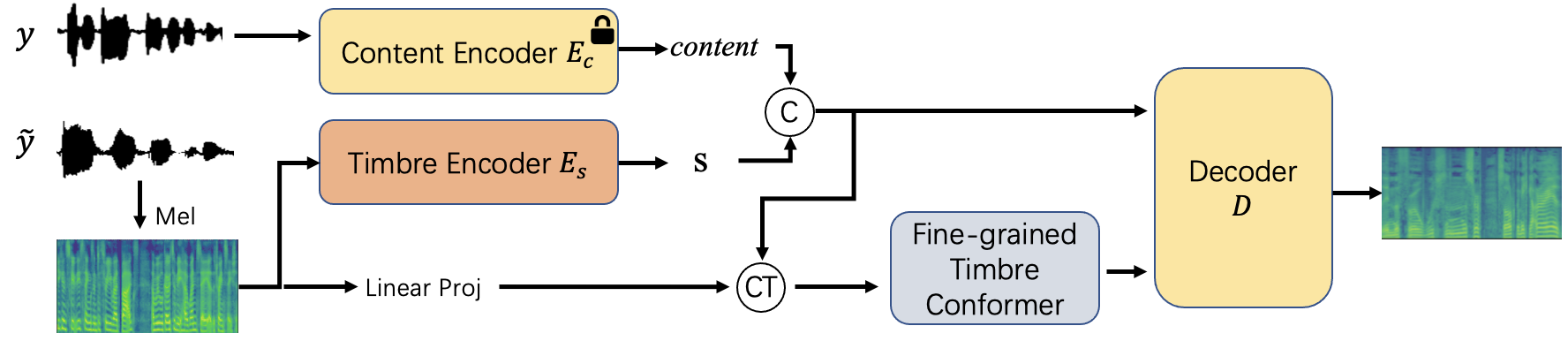}
    \caption{Model architecture of MulliVC. Note that modules printed with a \textit{lock} are frozen when training. We use \textbf{\textcircled{\tiny{C}}},\textbf{\textcircled{\tiny{CT}}} to denote concatenate along the channel axis and concatenate along the time axis respectively} 
    \label{fig:model}
\end{figure*}

\begin{figure}[!t]
    \centering
    \includegraphics[width=0.3\textwidth]{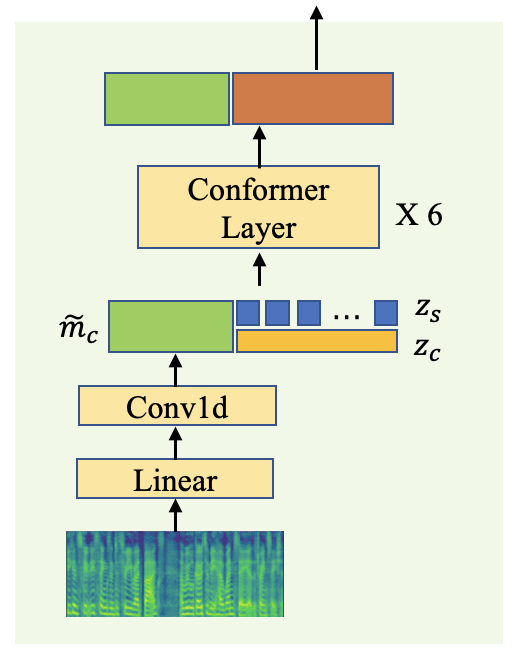}
    \caption{The Fine-grained Timbre Conformer architecture.} 
    \label{fig:conformer}
\end{figure}

\subsection{Cycle Consistency}
Due to the unavailability of speech data from the same speaker speaking two different languages, we addressed this limitation by simulating this scenario in step 3 through a cyclical approach encompassing steps 2 and 3.

In step 2, we employ speech\#3 of speaker 2 who speaks language B as content input and take speech\#2 of speaker 1 from language A as timbre input, and assume the model can generate a speech  SPK\_1|LAN\_B|\#3 which means the content is the same as speech\#3 but with the timbre of speaker 1. The loss of step 2 is
\begin{equation}
    \mathcal{L}=\lambda_1 \mathcal{L}_{Adv} + \lambda_3 \mathcal{L}_{Timbre}+ \lambda_5\mathcal{L}_{ASR}
\end{equation}
Where $\mathcal{L}_{ASR} = \lVert f_{A}(m_t) - f_{A}(\hat{m_t}) \rVert_2$ is the ASR perceptual loss. $\lambda_5=0.5$, $f_A$ is a pre-trained automatic speech recognition model, we use the last layer embedding of $f_{A}$ to calculate the perceptual loss. Considering there is no ground truth data of SPK\_1|LAN\_B|\#3, ASR perceptual loss is necessary to align the content information. Furthermore, the timbre loss ensures that the generated speech matches the timbre of speaker 1.

In step 3, we utilize the output SPK\_1|LAN\_B|\#3 obtained in step 2 as the timbre input. As for the content input, we use speech\#4 of speaker 1 speaking language A.  Since both speeches possess the timbre of speaker 1, we simulated the same speaker speaking two different languages. Consequently, another cross-lingual voice conversion can be performed. Moreover, we have the ground-truth data SPK\_1|LAN\_B|\#4 available during this step, which enables us to calculate the pitch perceptual loss and reconstruction loss. These losses ensure the model's output aligns with the ground-truth data distribution. The loss calculation in step 3 follows the same methodology as step 1.

By incorporating step 1 and the cross-lingual voice conversion cycle of step 2 and step 3, we guarantee the model's ability to convert voices within the same language while also enhancing the performance of cross-lingual voice conversion.

Additionally, the previous voice conversion scheme solely consisted of step 1, where the content features generated by SSL still contained certain timbre information. During audio reconstruction, the model unavoidably reads timbre information from the content features, resulting in insufficient disentanglement of timbre and content and sub-optimal generalization capabilities. Conversely, in our training strategy, step 2 ensures that the content and timbre inputs originate from different speakers, compelling the model to exclusively extract timbre information from the timbre inputs. This approach enhances the model's ability to disentangle timbre and content, thereby improving its generalization capabilities in terms of timbre.

\subsection{Model Architecture}
As illustrated in Figure \ref{fig:model}. Denote $Y=\{y_1,..., y_n\}$ as the speech corpus for a certain speaker. In training, $Y$ is partitioned as target audio $y_t$ and reference audio $\tilde{y}$. $y_t$ is fed into the pre-trained content encoder $E_c$ to get the content feature $z_c \in R^{T\times C} $, where $T$ is the length of time-axis and $C$ is channels. We adopt ContentVec\cite{contentvec} which aims to disentangle speaker information from audio and only encode content information as our content encoder.$\tilde{y}$ is fed into the global timbre encoder $E_s$ to encode the global timbre feature $S\in R^{1\times C}$. Inspired by MegaTTS2\cite{megatts2} and CDFSE\cite{cdfse} that fine-grained timbre information can represent the speakers' speaking habits and better help the model imitate the timbre of the reference audio, we design a Fine-grained Timbre Conformer\cite{conformer} to capture fine-grained timbre information. As illustrated in Figure ~\ref{fig:conformer}, the global timbre feature $S$ is first repeated along the time axis to $z_s\in R^{T\times C} $ and concatenated with $z_c$ in the channel axis to get feature $z_u\in R^{T\times 2C}$. The reference mel-spectrogram $\tilde{m}\in R^{T'\times D}$ where $D$ denotes the number of mel bins is firstly compressed into acoustic hidden states by a factor of $d$ in length then projected by a linear layer to  $\tilde{m_c}\in R^{\frac{T'}{d}\times 2C}$ and concatenate with $z_u$ along time axis sending to Conformer. In conformer, fine-grained timbre information will be merged with $z_u$ by Convolution and Self-Attention\cite{attention} mechanism.

\section{experiments}

\begin{table*}[!h]
    \centering
  \small
    \begin{tabular}{c|cccc|cccc|cccc}
    \toprule 
    & \multicolumn{4}{c|}{VCTK} & \multicolumn{4}{c}{VCTK-AIS1} & \multicolumn{4}{c}{AIS1-VCTK} \\
     Model  & nMOS↑    & sMOS↑    & WER↓   & SIM↑  & nMOS↑    & sMOS↑    & WER↓   & SIM↑ & nMOS↑    & sMOS↑    & CER↓   & SIM↑ \\
    \midrule
    \rowcolor{gray!25} GT  & -  & -  & 1.27  & - & - & - & 1.02 & - & - & - & 5.98 & - \\ 
    Diff-HierVC  & 3.44$\pm$0.11  & 3.41$\pm$0.09   & 6.46   & 0.276 & 3.34$\pm$0.13 & 3.47$\pm$0.07 & 5.96 & 0.184  & 3.17$\pm$0.12 & 3.27$\pm$0.12 & 26.99 & 0.180   \\
    FreeVC  & 3.72$\pm$0.09  &  3.63$\pm$0.08  &  3.10  & 0.376 & 3.79$\pm$0.08 & 3.71$\pm$0.09 & 2.88 & 0.143 & 3.51$\pm$0.10 & 3.43$\pm$0.08 & 22.62 & 0.299 \\
    FreeVC*  & 3.48$\pm$0.09  &  3.51$\pm$0.08  &  5.57  & 0.220 & 3.62$\pm$0.10 & 3.66$\pm$0.07 & 6.37 & 0.177  & 3.34$\pm$0.11 & 3.37$\pm$0.09 & 15.56 & 0.145 \\
    ConsistencyVC  & 3.78$\pm$0.10 &  3.59$\pm$0.09  & 4.86 & 0.174 & 3.92$\pm$0.08 & 3.88$\pm$0.08 & 5.48 & 0.256 & 3.66$\pm$0.11 & 3.52$\pm$0.10 & \textbf{9.58} & 0.094 \\
    Ours   & \textbf{3.92$\pm$0.11} & \textbf{3.88$\pm$0.11} & \textbf{2.24} & \textbf{0.395} & \textbf{4.00$\pm$0.08} & \textbf{3.98$\pm$0.09} & \textbf{2.37} & \textbf{0.376} & \textbf{3.69$\pm$0.11} & \textbf{3.73$\pm$0.09} & 9.91 & \textbf{0.311} \\
    \bottomrule
    \end{tabular}%
    \caption{The zero-shot voice conversion performance comparison of our model and baselines. "-" means the result is not available.}
    \label{tab:main}%
\end{table*}%

\begin{table*}[!h]
    \centering
    \begin{tabular}{c|cccc|cccc}
    \toprule 
    & \multicolumn{4}{c|}{EMIME Eng-Man} & \multicolumn{4}{c}{EMIME Man-Eng} \\
     Model  & nMOS↑    & sMOS↑    & WER↓   & SIM↑  & nMOS↑    & sMOS↑    & CER↓   & SIM↑  \\
    \midrule
    \rowcolor{gray!25} GT  & -  & -  & 4.29  & - & - & - & 2.97 & -  \\ 
    Diff-HierVC  & 3.39$\pm$0.08  & 3.41$\pm$0.08  & 8.45 & 0.426 & 3.42$\pm$0.08 & 3.46$\pm$0.06 & 6.84 & 0.395 \\
    FreeVC  & 3.64$\pm$0.06  & 3.54$\pm$0.07  & 7.72 & 0.331 & 3.56$\pm$0.05 & 3.50$\pm$0.08 & 6.70 & 0.309  \\
    FreeVC*  & 3.59$\pm$0.06  & 3.61$\pm$0.07  & 7.99 & 0.363 & 3.52$\pm$0.07 & 3.57$\pm$0.07 & 7.57 & 0.380  \\
    ConsistencyVC  & 3.83$\pm$0.08 & 3.72$\pm$0.07 & 5.28 & 0.322 & 3.80$\pm$0.06 & 3.71$\pm$0.06 & 8.14 &  0.310 \\
    Ours  & \textbf{4.02$\pm$0.08} & \textbf{4.00$\pm$0.07} & \textbf{5.21} & \textbf{0.534} & \textbf{3.96$\pm$0.10} & \textbf{4.03$\pm$0.08}  & \textbf{4.49} & \textbf{0.549}  \\
    \bottomrule
    \end{tabular}%
    \caption{Zero-shot voice conversion performance comparison of our model and baselines on EMIME bilingual dataset. EMIME Eng-Man means the source speech records come from English speech corpus and the targets are from Chinese Mandarin speech corpus. EMIME Man-Eng is vice versa.}
    \label{tab:emime}%
\end{table*}%

\subsection{Experimental Setup}\label{4.1}
\subsubsection{Dataset} \label{data}
\
\newline
We use several datasets to train our model. We use Libritts~\cite{libritts} for the English speech corpus. And we use aidatatang\_200zh\footnote{\url{https://openslr.org/62/}}, MAGICDATA\footnote{\url{https://openslr.org/68/}} and ST-CMDS\footnote{\url{https://openslr.org/38/}} Chinese Mandarin speech corpus. We use VCTK\cite{vctk}, Aishell-1\cite{aishell1} and EMIME\cite{emime} datasets to evaluate our models' zero-shot voice conversion performance. EMIME\cite{emime} contains bilingual audio recordings by the same speakers. The sample rate is 16KHz for all speech data.

\subsubsection{Model Configuration} 
\
\newline
MulliVC consists of a content encoder, a timbre encoder, a fine-grained timbre conformer, a mel decoder, and a Patch-GAN discriminator. The timbre encoder consists of 5 convolution layers with 512 hidden size and 5 kernel size. The Fine-grained timbre conformer consists of 6 conformer layers. The mel decoder consists of 5 convolutional blocks with 512 hidden size and 5 kernel size. In the training stage, we involve three pre-trained models to calculate auxiliary loss: a speaker verification model, an automatic speech recognition model, and a pitch predictor. These models are trained on the same dataset of MulliVC, please refer to Appendix A for the details of these models.

\subsubsection{Training Details}
\
\newline
MulliVC is trained on 1 A100 GPU with a batch size of 8 speeches. Considering 1 training step is split into 3 substeps, the model takes 24 speeches as input for 1 step in total. We use the Adam optimizer with learning rate of  $10^{-4}, \beta{_1} = 0.9, \beta{_2} = 0.999, \epsilon = 10^{-9}$. We train MulliVC for 240K training steps. In training, language A and language B will be randomly switched. The predicted mel-spectrograms are transformed into audio samples using pre-trained HiFi-GAN V1~\cite{hifigan}.

\subsubsection{Evaluation Metrics}
\
\newline
Following \cite{diffhiervc}, we conduct the naturalness and similarity mean opinion score (nMOS and sMOS, respectively) for subjective evaluation, 16 subjects are employed to provide the subjective measures. we evaluate the word error rate (WER) of the English corpus, the character error rate (CER) of the Chinese corpus, and speaker similarity (SIM) for objective evaluation. We use whisper-large-v3\cite{whisper} to transcribe the generated speech into text. Then, the WER/CER between the transcribed text and the original target text is measured. In terms of the cosine speaker similarity, we use the WavLM-TDCNN model\footnote{\url{https://github.com/microsoft/UniSpeech/tree/main/downstreams/speaker_verification}}\cite{wavlm-tcdnn} to compute the cosine speaker similarity score between the target speech and the converted speech. The similarity score is in the range of $[-1, 1]$, where a larger value indicates a higher similarity of input samples.

\subsection{Main Results}

\subsubsection{Baseline Models}
\
\newline
We use pre-trained FreeVC\footnote{\url{https://github.com/OlaWod/FreeVC}}, ConsistencyVC\footnote{\url{https://github.com/ConsistencyVC/ConsistencyVC-voive-conversion}} and DiffHier-VC\footnote{\url{https://github.com/hayeong0/Diff-HierVC}} as our baseline models. 
The pre-trained FreeVC was trained on the VCTK dataset, which is not the zero-shot scenario. To train with the same settings as our model, we retrain FreeVC with our training dataset and denote the retrained model as FreeVC*. DiffHier-VC is trained on the LibriTTS dataset. ConsistencyVC is trained on a combination of English, Chinese, and Japanese datasets.

\subsubsection{Zero-shot Voice Conversion Comparison}
\
\newline
We conducted subjective and objective evaluations for three zero-shot voice conversion (VC) scenarios. The first scenario involved voice conversion in English, using source and target speeches from the VCTK dataset. Additionally, we conducted two cross-lingual voice conversion experiments: VCTK-AIS1 and AIS1-VCTK. This implies using source speeches from VCTK and target speeches from Aishell-1 for the former experiment, and source speeches from Aishell-1 and target speeches from VCTK for the latter. For objective evaluation of each experiment, we randomly created 400 speaker pairs, and each pair was randomized to use 5 speeches, for a total of 2000 speech pairs to calculate WER and SIM. For human evaluation of each experiment, we randomly select 30 synthesized speech records to conduct nMOS and sMOS evaluation. The results are listed in Table ~\ref{tab:main}.

For speech intelligibility, our method achieves lower WER compared with baseline models in VCTK and VCTK-AIS1. The CER metric of our method on the AIS1-VCTK dataset is comparable to ConsistencyVC. Also, we achieved the highest nMOS score on all datasets. 
For speaker similarity, the SIM score and sMOS score of our method are significantly improved compared to the baselines. It is worth noting that FreeVC is trained on the VCTK dataset, while our method surpasses FreeVC on the VCTK dataset, indicating that the zero-shot performance of our method outperforms the performance of FreeVC's seen speaker. In addition we observed that ConsistencyVC had lower SIM scores than FreeVC* and DiffHierVC under VCTK and AIS1-VCTK tests, but obtained higher sMOS scores. It suggests that WavLM-TDCNN pays attention to some details that are weaker in human perception compared to human rators. In addition, the clarity and naturalness of the audio also affect the sMOS scores compared to WavLM-TDCNN. Speeches with low intelligibility may also receive high SIM scores, as further confirmed by our research in the ablation study section.

We further compare our model's zero-shot voice conversion performance with baselines on the EMIME bilingual dataset, the results are displayed in Table ~\ref{tab:emime}. On the EMIME dataset, the results of our model have a significant advantage over the baselines. In addition, we find that the SIM scores of the models are generally higher than those of AIS1-VCTK and VCTK-AIS1 because the bilingual speakers of EMIME are native speakers of Chinese, so they have similar accents when speaking the two languages. Higher SIM scores are obtained when the generated speech is similar in accent to the target speech.

To test the voice conversion capability on unseen languages, we test on the French (FR) and German (DE) subsets of M-AILABS~\cite{mailabs}. The results are listed in Table ~\ref{tab:unseen}. Our model is substantially ahead of the baseline model in unseen languages, meaning that training with cycle consistency enables the model to disentangle timbre for unseen languages and enhance the generalization capability. 

\begin{table*}[!h]
    \centering
    \begin{tabular}{c|cccc|cccc}
    \toprule 
    & \multicolumn{4}{c|}{FR-DE} & \multicolumn{4}{c}{DE-FR} \\
     Model  & nMOS↑    & sMOS↑    & WER↓   & SIM↑  & nMOS↑    & sMOS↑    & CER↓   & SIM↑  \\
    \midrule
    \rowcolor{gray!25} GT  & -  & -  & 5.80  & - & - & - & 5.50 & -  \\ 
    Diff-HierVC  & 3.44$\pm$0.09  & 3.52$\pm$0.11   & 14.78   & 0.260 & 3.55$\pm$0.11  & 3.55$\pm$0.08   & 12.09   &  0.212 \\
    FreeVC  & 3.76$\pm$0.11  &  3.61$\pm$0.09  &  13.00  & 0.179 & 3.73$\pm$0.09  &  3.59$\pm$0.08  &  10.18  & 0.106  \\
    FreeVC*  & 3.51$\pm$0.13 & 3.46$\pm$0.09 & 20.64 & 0.191 & 3.59$\pm$0.11 & 3.48$\pm$0.12 & 19.18 &  0.153 \\
    ConsistencyVC  & 3.62$\pm$0.09 &  3.59$\pm$0.08  & 19.48 & 0.149 & 3.69$\pm$0.08 &  3.61$\pm$0.11  & 13.06 & 0.111\\
    Ours & \textbf{3.86$\pm$0.10} & \textbf{3.82$\pm$0.09} & \textbf{7.34} & \textbf{0.410} & \textbf{3.85$\pm$0.08} & \textbf{3.78$\pm$0.10} & \textbf{6.63} & \textbf{0.332}  \\
    \bottomrule
    \end{tabular}%
    \caption{Zero-shot voice conversion performance comparison for unseen languages on the French (FR) and German (DE) subsets of  M-AILABS~\cite{mailabs} dataset. FR-DE means the source speech records come from French speech corpus and the targets are from German. DE-FR is vice versa.}
    \label{tab:unseen}%
\end{table*}%

\subsection{Method Analyses}

\subsubsection{Validating The Effectiveness Of SV Model for Cross-linguistic Scenario}
\
\newline
\begin{figure}[!t]
    \centering
    \includegraphics[width=0.5\textwidth]{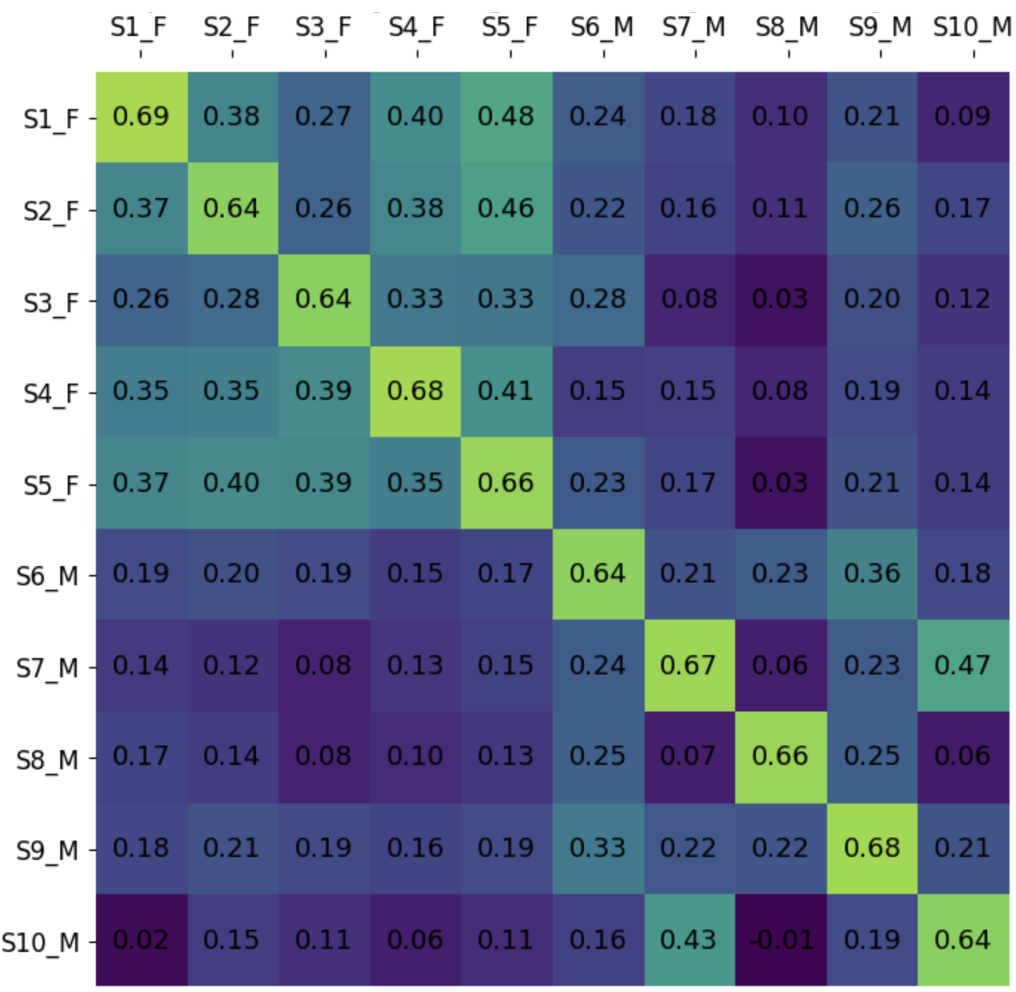}
    \caption{Si\_F, Sj\_M denotes female speaker i and male speaker j respectively. Speeches of Chinese Mandarin spoken -by the corresponding speaker are displayed on the horizontal axis. Speeches of English are displayed on the vertical axis. The number in each grid is the average SIM between the speeches.} 
    \label{fig:emime_compare}
\end{figure}

The SV model WavLM-TCDNN is trained using a multilingual dataset composed of monolingual speakers. Consider that we need to test SIM scores with the it to evaluate the VC model's ability to convert voice across languages. It is essential to determine whether the SV model can correctly identify that two speeches in different languages come from one speaker. We experiment on the EMIME\cite{emime} dataset which contains bilingual audio recordings by the same speakers. We sample 5 female speakers and 5 male speakers from the EMIME dataset, for each speaker, we sample 20 English speeches and 20 Chinese Mandarin speeches. And we calculate the average SIM with WavLM-TDCNN between speeches. The result is shown in Figure \ref{fig:emime_compare}. The consequent findings reveal that the similarity between speeches of different languages delivered by the same speaker is significantly higher than that between speeches of different languages delivered by different speakers. This pertinent evidence leads us to conclude that WavLM-TCDNN proves effective in measuring timbre similarity within cross-linguistic scenarios.

\subsubsection{Speaker Clustering Comparison}
\
\newline

\begin{figure*}[!t]
    \centering
    \includegraphics[width=\textwidth]{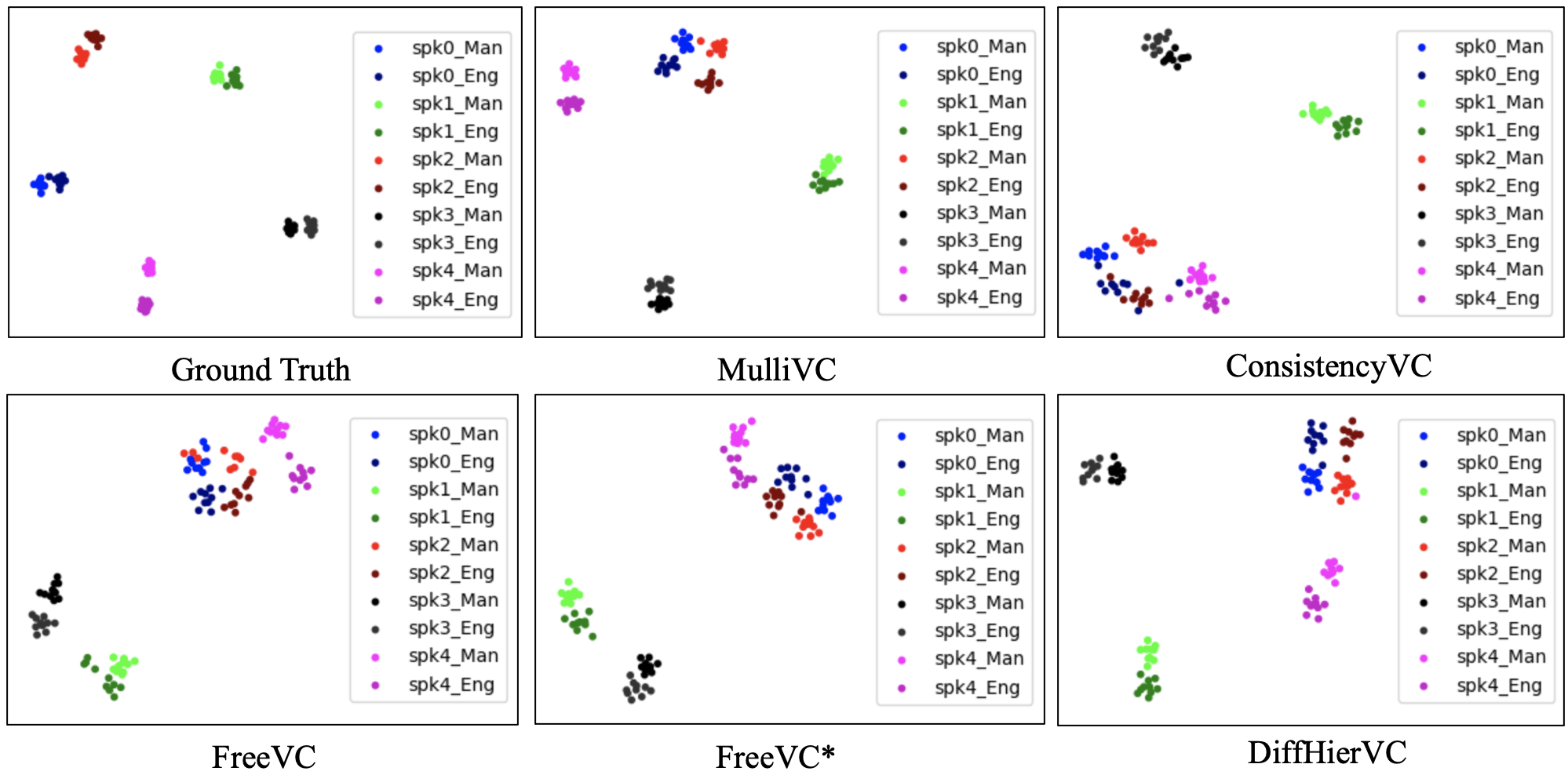}
    \caption{Visualization of the speaker embedding space based on t-SNE and four randomly selected speakers in the EMIME dataset. spk$i$\_MAN denotes the Chinese Mandarin speech of speaker $i$. And spk$i$\_ENG denotes the English speech of speaker $i$.}  
    \label{fig:tsne}
\end{figure*}
To further investigate the speaker embedding space of the WavLM-TDCNN models and explore the cross-lingual voice conversion performance of each model, we conducted an experiment on the EMIME dataset. We reconstructed language A/B by utilizing speech from language A/B as the content input and the same speaker's speech from language B/A as the timbre input. Utilizing WavLM-TDCNN, we obtained the reconstructed voice speaker embedding, which was then subjected to clustering using t-SNE~\cite{tsne}. The clustering results, visualized in Figure ~\ref{fig:tsne}, indicate that there is a minor difference between the distributions of speaker embeddings derived from speech in different languages spoken by the same individual. However, this difference is significantly smaller compared to the disparity observed between speaker embeddings from different speakers. This finding further validates Section 4.2's assertion regarding the applicability of WavLM-TDCNN in evaluating voice conversion within cross-language scenarios. Moreover, when comparing the distributions of speaker embeddings among different speakers, MulliVC exhibits more distinct and tightly grouped clusters, suggesting superior performance in timbre disentangling and voice conversion than the other baselines.

\begin{table*}[!h]
    \centering
    \begin{tabular}{cc|cc|cc|cc}
    \toprule 
    \multirow{2}{*}{Setting} & \multirow{2}{*}{Model} & \multicolumn{2}{c|}{VCTK} & \multicolumn{2}{c}{VCTK-AIS1} & \multicolumn{2}{c}{AIS1-VCTK} \\
     & &  WER↓   & SIM↑  & WER↓   & SIM↑   & CER↓   & SIM↑ \\
    \midrule
    \#1 & Ours & \textbf{2.24} & \textbf{0.395}   & \underline{2.37} & \underline{0.376}  & \underline{9.91} & 0.311 \\ 
    \midrule
    \#2 & w/o step 3  & 8.42  & 0.393 & 4.76 & 0.422  & 10.65 & \textbf{0.333} \\
    \#3 & w/o step 2,3   & 2.24   & 0.310 & \textbf{2.09} & 0.259 & \textbf{9.37} & 0.191 \\
    \#4 & w/o ASR loss   & 10.40  & 0.384 & 19.03 & \textbf{0.461} & 26.47 & \underline{0.319} \\
    \#5 & w/o Fine-Grained Confromer  & 6.96 & 0.337 & 5.45 & 0.363 & 30.47 & 0.265\\
    \bottomrule
    \end{tabular}%
    \caption{The ablation study of MulliVC. The design of our MulliVC achieves a favorable balance between intelligibility (WER) and speaker similarity (SIM).}
    \label{tab:ablation}%
\end{table*}%
\subsection{Ablation Study}

\subsubsection{Cross-lingual Steps}
\
\newline
To verify the effectiveness of the cross-lingual steps (step 2 and step 3), we evaluate the performance of the voice conversion models without them and list the results in Setting \#2 and Setting \#3 of Table~\ref{tab:ablation}.
Compare Setting \#2 with Setting \#3 of Table~\ref{tab:ablation}, speaker similarity decreases significantly after removing step 2. 
It is worth noting that after adding the cross-lingual voice conversion step 2, the speaker similarity of the VCTK dataset with timbre migration within the same language is also highly improved. This is because the embedding encoded by the content encoder holds part of the timbre information\cite{contentvec,nansy} when there is only intra-language timbre migration, the model tends to partly use the timbre information from the content encoder, resulting in insufficient timbre disentanglement. The cross-language timbre migration scenario of step 2 forces the model to encode timbre only from the timbre input, which improves the timbre disentanglement ability of the model. On the other hand, adding step 2 leads to a rise in WER and CER, further addition of step 3 makes WER and CER similar to that of only step 1, which shows that ASR loss is not enough to align the content, the reconstruction loss in step 3 is important. 

\subsubsection{ASR Perceptual Loss} 
\
\newline
We evaluated the performance of the model when removing asr perceptual loss from step 2, and removing asr perceptual loss leads to an increase in WER and CER for all three scenarios. However, adding asr perceptual loss at the same time leads to a decrease in the speaker similarity, because although the asr model's training target is CTC loss, even though the last layer of the output embedding still inevitably encodes some of the timbre information. When we optimize the pairing of SPK\_1|LAN\_B|\#3 and SPK\_2|LAN\_B|\#3 in step 2 will negatively affect the timbre disentanglement.

\subsubsection{Fine-grained Timbre Conformer} 
\
\newline
Removing the Fine-grained timbre conformer leads to a double decrease in the intelligibility of the generated speech and speaker similarity. fine-grained timbre conformer facilitates the interaction between fine-grained timbre and content information, with positive effects on both WER and SIM.

\section{Conclusion}
This research paper presents MulliVC, a multi-lingual VC system designed for high-fidelity timbre migration and mel-spectrogram generation. The proposed three-step training architecture enhances the model's performance in speaker adaptation, both within and across languages. And the Fine-grained timbre conformer component improves the speaker similarity and intelligibility of the generated speech. The experimental results demonstrate that our model surpasses the state-of-the-art in both intra-language and cross-lingual zero-shot voice conversion scenarios.

However, despite the considerable improvement in speaker adaptation achieved by our method, several aspects still require further enhancement, particularly in zero-shot scenarios. Firstly, the current training dataset utilized in our experiments remains relatively small. Consequently, it may not be adequate for tasks such as movie dubbing, where expressive voices and highly diverse timbre characteristics are prevalent. Additionally, the employed content encoder retains some prosody and timbre information, which hinders the effective separation of timbre from content. Moreover, the computation of timbre loss relies on a pre-trained speaker verification (SV) model, which could benefit from larger datasets encompassing more languages to enhance its accuracy. This, in turn, would contribute to better speaker adaptation.


\bibliographystyle{ACM-Reference-Format}
\bibliography{sample-base}

\appendix
\begin{figure}[!ht]
    \centering
    \includegraphics[width=0.3\textwidth]{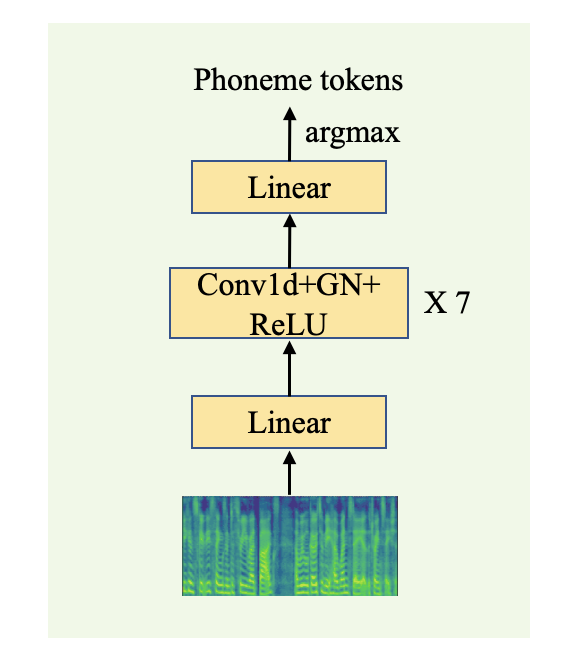}
    \caption{The ASR model architecture.} 
    \label{fig:asr}
\end{figure}

\section{Detailed Experimental Settings}\label{app:train_details}

\subsection{Speaker Verification Model}\label{app:sv}
In this section, we introduce our pre-trained speaker verification model which takes mel-spectrogram as input to calculate timbre loss. The model's architecture is the same as the speaker encoder described in section~\ref{4.1}, with a linear layer in the last to project the embedding from 512 to 256. The model is trained by distillation, with WavLM-TDCNN as the teacher model and the MSE loss between WavLM-TDCNN's output and our model's output. The model is trained by 240K timesteps on our train set, with batchsize=48.

\subsection{ASR model}\label{app:asr}

Here we introduce our pretrained ASR model to calculate ASR loss. Our asr model takes the mel-spectrogram as input and predicts the phoneme corresponding to each of the 4 melbins. The phoneme is obtained and aligned with speech by external alignment tools MFA~\cite{mfa}. The model was trained using CTC loss with 160K time steps on the training set and batch size=48. The architecture of ASR model is displayed in Figure~\ref{fig:asr}

\subsection{Pitch Predictor}\label{app:pitch}
We use REAPER~\cite{reaper} to extract F0(pitch) from raw audio, and interpolate the F0's length with mel-spectrogram. We train the pitch predictor to predict F0 from mel-spectrogram and calculate MSE loss with the extracted F0. We adopt the pitch predictor architecture from FastSpeech2~\cite{fastspeech2} as the architecture of our pitch predictor.

\end{document}